\documentclass[runningheads]{llncs}

\usepackage[T1]{fontenc}
\usepackage{amsmath}

\usepackage{graphicx}

\usepackage[hidelinks]{hyperref}
\hypersetup{pdfauthor={},pdftitle={Title},pdfcreator={},pdfproducer={}}

\usepackage{todonotes}
\usepackage{booktabs}
\usepackage[font=small,labelfont=bf]{caption}
\usepackage{subcaption}
\usepackage{algorithm}
\usepackage{algorithmic}

\begin{document}

\title{The Curious Case of High-Dimensional Indexing as a File Structure: A Case Study of eCP-FS}

\titlerunning{The Curious Case of High-Dimensional Indexing as a File Structure}

\author{Anonymous Authors}
\author{
    Omar Shahbaz Khan\inst{1} \orcidID{0000-0001-9720-3645} \\ 
    Gylfi {\TH}\'{o}r Gu\dh{}mundsson\inst{1} \orcidID{0000-0003-0846-6617} \\ 
    Björn {\TH}\'{o}r J\'{o}nsson\inst{1} \orcidID{0000-0003-0889-3491}
}

\authorrunning{O.S. Khan, G.{\TH}{}. Gu\dh{}mundsson and B.{\TH}{}. J\'{o}nsson}

\institute{Reykjavik University, Reykjavík, Iceland \\
\email{\{omark, bjorn, gylfig\}@ru.is}}

\maketitle              

\newcommand{\ok}[1]{\todo[inline,color=pink!40,author=Omar]{ #1}}
\newcommand{\gy}[1]{\todo[inline,color=orange!40,author=Gylfi]{ #1}}
\newcommand{\bj}[1]{\todo[inline,color=green!40,author=Björn]{ #1}}

\newcommand{\exq}[0]{Exquisitor}
\newcommand{\ecpfs}[0]{eCP-FS}

\begin{abstract}
Modern analytical pipelines routinely deploy multiple deep learning and retrieval models that rely on approximate nearest-neighbor (ANN) indexes to support efficient similarity-based search. While many state-of-the-art ANN-indexes are memory-based (e.g., HNSW and IVF), using multiple ANN indexes creates a competition for limited GPU/CPU memory resources, which in turn necessitates disk-based index structures (e.g., DiskANN or eCP).  In typical index implementations, the main component is a complex data structure that is serialized to disk and is read either fully at startup time, for memory-based indexes, or incrementally at query time, for disk-based indexes.  To visualize the index structure, or analyze its quality, complex coding is needed that is either embedded in the index implementation or replicates the code that reads the data structure. In this paper, we consider an alternative approach that maps the data structure to a file structure, using a file library, making the index easily readable for any programming language and even human-readable. The disadvantage is that the serialized index is verbose, leading to overhead of searching through the index. The question addressed in this paper is how severe this performance penalty is. To that end, this paper presents eCP-FS, a file-based implementation of eCP, a well-known disk-based ANN index. A comparison with state-of-the-art indexes shows that while eCP-FS is slower, the implementation is nevertheless somewhat competitive even when memory is not constrained. In a memory-constrained scenario, eCP-FS offers a minimal memory footprint, making it ideal for resource-constrained or multi-index environments.

\keywords{High-Dimensional Indexing \and Resource Constrained Search \and Incremental Retrieval \and Disk-based ANN}
\end{abstract}

\section{Introduction}

To gain insight from large-scale media collections, a mix of exploratory and search-oriented analytics methods---such as clustering, similarity search, interactive browsing, and LLM or VLM summarization---are employed.
For retrieval oriented tasks, it is common to utilize Approximate Nearest Neighbor (ANN) search guided by indexes which organize high-dimensional feature vectors, such as text embeddings or image descriptors, to enable rapid retrieval of items most similar to a given query. 
Depending on hardware constraints and performance requirements, these indexes (and their data) can reside entirely in memory, HNSW~\cite{hnsw}, IVF~\cite{ivf} 
or Annoy~\cite{annoy}, or be disk-based, such as DiskANN~\cite{diskann-github} or eCP~\cite{ecp}. 
Being disk-based has the advantage of allowing the index structure to choose between memory and disk, minimizing memory footprint as needed. 

Modern analytical systems rarely stop at retrieval from a single index. It is common to run multiple resource intensive methods in parallel, like deploying several deep-learning models/LLMs/VLMs for inference, some residing in GPU VRAM while others reside in RAM. This is taxing on laptops/desktops or even modest servers.
Furthermore, if the analytical system is not running on a dedicated server, other unrelated processes may also demand resources.
In-memory ANN indexes keep the footprint low by using compression (product quantization) or dimensionality reduction (PCA) at the risk of sacrificing some accuracy.
By contrast, disk-based algorithms do not need to risk losing accuracy as they can preserve the original data, unaltered, on disk.
In such analytical scenarios, choices must thus be taken such as the specific index type (clustering-based, graph-based, hierarchical), whether or not to use compression or dimensionality reduction, and whether it needs to be an in-memory or disk-based index. 

State-of-the-art ANN indexes are typically highly optimized, and either provided as black-box software or unintentionally opaque, hiding how they work behind complex code. This results in valuable information, such as their structure, being inaccessible for further analysis and information mining.
Were it possible, we believe that interacting with the underlying index structure could prove insightful for some analytical tasks, and also assist in discovering issues with data collections, such as skewed data distributions.

We present {\bf eCP-FS}, an ANN indexing algorithm implemented as a file system. It is based on eCP but unlike its predecessor it fully exposes its internal structure in a easy and relatable way (through files and folders) while minimizing its in-memory overhead in a tunable way.
By leveraging the Zarr storage format, eCP-FS offers language-agnostic access across diverse programming environments, and it is designed for extensibility, allowing embedded metadata and alternative vector representations. 
Furthermore, we extend the eCP search algorithm with a focus on incremental retrieval, a capability absent from existing ANN libraries. 
This is, as we  show, an incredibly useful addition in long running interactive retrieval scenarios, commonly found in exploration and/or browsing of large datasets. 
We compare eCP-FS against leading in-memory and disk-based ANN indexes with a new automated multimedia retrieval benchmark inspired by interactive live-search competitions. 
As expected, eCP-FS matches the characteristic accuracy profile of eCP but incurs higher latency due to disk I/O. Its real advantages, however, lie in incremental search performance, transparent data structures, and a minimal memory footprint---making eCP-FS especially well suited to resource-constrained machines, hybrid analytical pipelines that combine heavy models, or systems managing multiple co-located indexes.
\section{Background}

Indexing large scale collections of high-dimensional data is commonly achieved through approximate nearest neighbor~(ANN) indexes. In this paper we focus on the state-of-the-art cluster- and graph-based algorithms. 

{\bf IVF}, or Inverted File, is clustering-based technique that has become a stable of information retrieval. A popular implementation of it can be found in Facebook's FAISS library~\cite{ivf}. Features are extracted from documents and clustered, typically using k-Means. Each centroid then keeps track of the file IDs its features came from. At search-time, \texttt{nprobe} is used to determine the search expansion (i.e.~how many clusters to check) and Term Frequency-Inverse Document Frequency (TF-IDF) is used to calculate document similarity.

{\bf HNSW} or Hierarchical Navigable Small World is a state-of-the-art algorithm that delivers good accuracy at reasonable resource use. At search-time, the layers of graphs are used to quickly navigate to a good starting point, fallowed by a greed search over the bottom layer graph k most similar items. The main down side of HNSW is the cost of building the index structure and the lack of scalability at construction time (full dataset must fit in RAM).

{\bf DiskANN} is a graph-based algorithm from Microsoft, designed for indexing and searching  datasets that are too large to fit in main memory. This is achieved by storing its data (index and vectors) on secondary storage such as SSD. 
Unfortunately, and somewhat in contradiction to its intended use, each index structure can only be built in-memory. However, once written to disk, multiple such indexes can be merged into one structure that can then be partially loaded as needed at search-time.

\section{The eCP Index}

eCP is a hierarchical cluster-based index, where the cluster representatives (or leaders) are randomly selected from the indexed collection. While crude, it is simple and fast. More complex calculation of the cluster leaders is of course possible (such as full K-Means etc.).
The key aspect of eCP is that it  builds the index top-down, which allows using the index structure to speed up the indexing process. The index takes three parameters, the desired cluster size $C$ (based on the optimal page size for the underlying storage hardware), the size of the feature vector $V$ and the index depth $L$.  From those parameters, the number of cluster leaders $l$ is computed as 
$
l = N \cdot V/ C
$.
Index traversal cost over the $L$ deep index is minimized by creating $w = \sqrt[\leftroot{5} \uproot{3} L]{l}$ internal subsets per level, resulting in $L\cdot w$ distance calculations if only the single most promising (most similar) subbranch is followed.

This is best explaned with an
example: Consider $N$=$1\text{ million}$ CLIP embeddings of $1152$ dimensions using $\text{float16}$, for $V = 2 \cdot 1152 = 2304$, a desired cluster size on disk of $C=128 \mbox{KB} = 131 072$, and an index depth of $L=3$.  This gives $V / N = 131072 / 2304 = 57$ as the desired cluster size, $l = 1 000 000 / 57 \approx 17 544$ leaders, and the optimal internal subset for this $L$=3 level deep index is 
$\sqrt[\leftroot{5} \uproot{3} 3]{17544} \approx 26$
The top level of the index thus has 26 nodes, each linking to 26 second level nodes, which  in turn link to 26 leaf nodes (clusters), of approximately 57 descriptors each.
Query cost, in its simplest form, is thus $3 \cdot 26 +57  = 135$ distance calculations on average.\footnote{There is no guarantee that each cluster created this way will have exactly $C / V$ descriptors, as that is dependent on the internal data representations of the collection being indexed. If, for example, there are 100K identical images they will end up in the same cluster, as each image (descriptor) will be inserted top down.}




The index traversal of single most promising edge, described above, is naive. In reality a more complex $k$-NN search is used, involving either {\it search expansion} or {\it soft assignments}.
We will focus only on expanding the search here, namely that instead of following only the single most promising (similar) edge between index levels, we expand the search to the $b$ most similar edges, i.e.~$b$ is the expansion parameter.
The query cost for the expanded search will thus be $w +  + (L-1)\cdot b\cdot w + b \cdot N/V$ on average.
\section{The Hierarchical Index as File Structure}

\subsection{Index File Structure}

\begin{figure}[t!]
    \centering
    \includegraphics[width=0.35\linewidth]{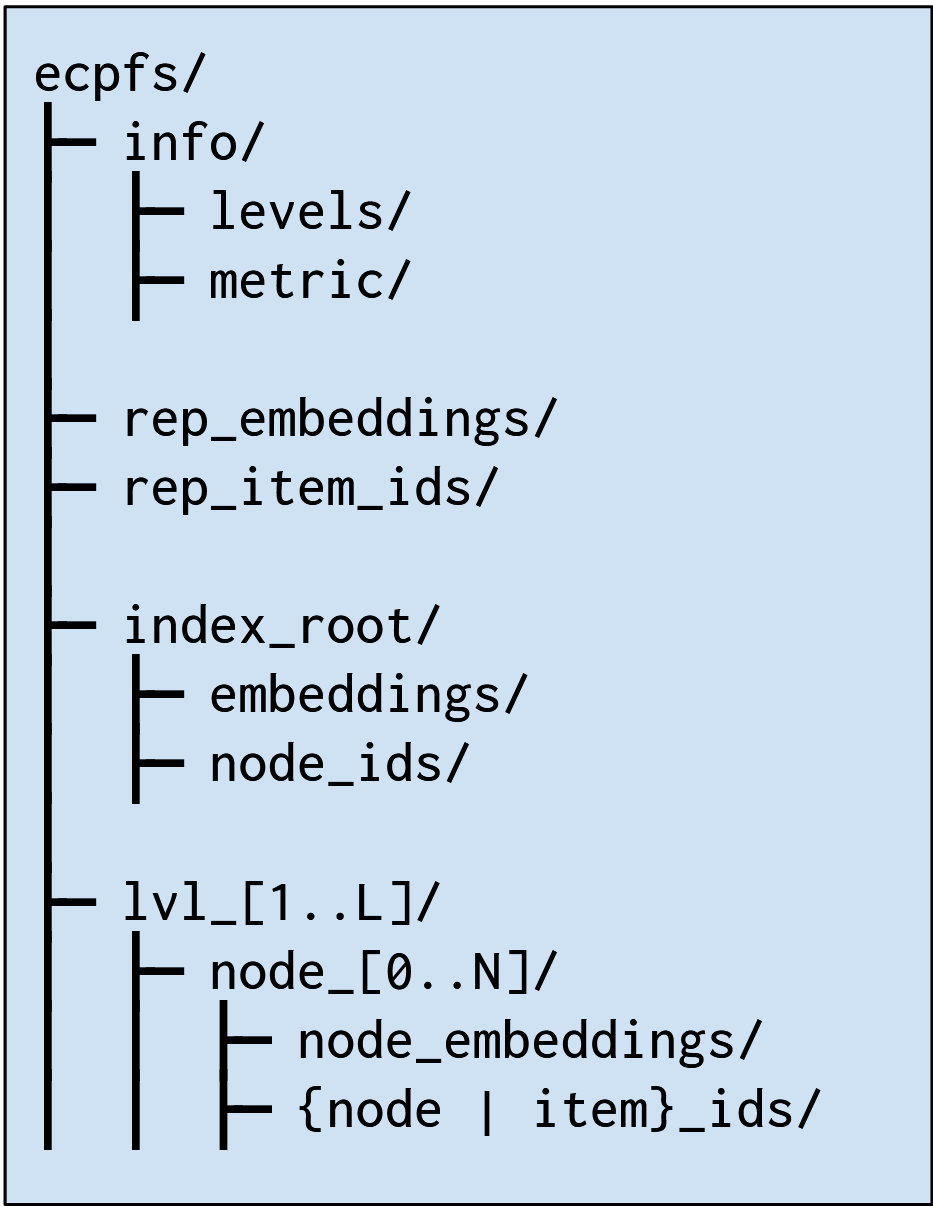}
    \caption{eCP-FS layout}
    \label{fig:ecpfs:layout}
\end{figure}

Figure~\ref{fig:ecpfs:layout} depicts the layout of the eCP index as a file structure, which we refer to as eCP-FS. The \texttt{info} group contains basic index information such as the maximum level of the index and the metric used for building the index. The \texttt{rep\_embeddings} and \texttt{rep\_item\_ids} are the embeddings and ids of the items selected as representatives to construct the index top-down. The \texttt{index\_root} group contains the embeddings and ids to the first level centroids. \texttt{lvl\_[0..L]} are the groups containing the nodes for each level, where a node group consists of embeddings and ids to the next level similar to the index root.
This is the base layout of the index, which can be extended to include data for any number of features. To ensure longevity, support, and language agnostic flexibility, and concurrency, eCP-FS has been implemented using the Zarr library.

\subsection{Retrieval}
Assuming the eCP-FS index has been built and is ready to be used for retrieval, the first step in loading the index is to read the \texttt{info} and \texttt{index\_root} group, followed by constructing node objects for each level. At this point, however, no data from the nodes is read. When nodes are accessed through search queries, their data is kept in memory, which leads to faster retrieval times as the index is used and therefore read into memory.
Taking inspiration from caching mechanisms, we can have strict control over the memory footprint. We set a limit (upper bound) on how many nodes can be in RAM at any given time, paired with a Least Recently Used (LRU) policy to free up space as needed. This is both flexible and tunable as the limit and the replacement policies could be changed at run-time.
Furthermore, pre-fetching options can be applied to fetch nodes up to a specific level, using background threads.

\subsection{Incremental Retrieval}
Contrary to the original eCP search methodology of finding the $b$ best nodes from a level before continuing to the next level, eCP-FS uses a priority queue $T$ and opens the best node regardless of level, adding its children to $T$.
The benefit of having a single priority queue $T$ and item vector $I$, is the potential and ease of supporting incremental (or resuming a) search request. 
This can occur from requests originating internally or externally as follows:
\begin{itemize}
    \item \textbf{Internal:} 
    Many modern search engines allow users to define extra search constraints, in the form of filters.
    If the user has applied filters, it is entirely possible that some (or all) of the $k$ leaf items selected 
    fail to pass the filtering and the initial search falls short of $k$ items it should return. In this scenario, eCP-FS expands the $b$ setting by a factor of 2, and continues (resumes) the search.
    \item \textbf{External:} A query state is stored internally in $QS$, along with $T$ and $I$. After a search the user is given the result and a query id, which they can use to request more results for that query. This methodology supports multiple active search queries, leading to a flexible retrieval experience over repeated single shot queries with increased $k$.
\end{itemize}

\begin{algorithm}[t!]
\caption{Starting a new search}
\label{alg:new-search}
\begin{algorithmic}[1]
    \REQUIRE 
        $q$: input vector, 
        $k$: number of items to return, 
        $b$: search expansion parameter, 
        $mx\_inc$: number of times $b$ can expand,
        $E$: set of items to exclude / filter 
    \STATE \textbf{function} \textsc{NewSearch} ($q,\,k,\,b,\,mx\_inc,\,E$)
    \STATE \quad Let $Q \leftarrow \{\,q,\;T \leftarrow \emptyset,\;I \leftarrow \emptyset\,\}$
    \STATE \quad $QS \leftarrow Q$
    \STATE \quad Let $\mathit{q\_id} \leftarrow (\bigl|QS| - 1)$
    \STATE \quad \textsc{IncrementalSearch} $(q\_id,\,k,\,b,\,mx\_inc,\,E)$
    \STATE \quad $\mathit{results} \leftarrow$ \textsc{GetNextKItems} $(q\_id,\,k,\,b,\,max\_inc,\,E)$
    \STATE \quad \textbf{return} $(results,\,q\_id)$
\end{algorithmic}
\end{algorithm}

\begin{algorithm}[t!]
\caption{Get $k$ items from an active query}
\label{alg:get-next-k-items}
\begin{algorithmic}[1]
    \REQUIRE 
        $\mathit{q\_id}$, 
        $k$, 
        $b$, 
        $mx\_inc$,
        $E$ : See Algorithm~\ref{alg:new-search} for descriptions
    \ENSURE $q\_id$ is valid

    \STATE \textbf{function} \textsc{GetNextKItems}$(\mathit{q\_id},\,k,\,b,\,mx\_inc,\,E)$
    \STATE \quad Let $Q \leftarrow QS[\mathit{q\_id}]$
    \STATE \quad Let $\mathit{cnt} \leftarrow \min\bigl(\lvert Q.I\rvert,\;k\bigr)$
    \STATE \quad \textbf{if} $\mathit{cnt} = 0$ \AND $Q.T$ is not empty \textbf{then}
        \STATE \quad\quad \textsc{IncrementalSearch}$(\mathit{q\_id},\,k,\,b,\,mx\_inc,\,E)$
        \STATE \quad\quad Let $\mathit{cnt} \leftarrow \min\bigl(\lvert Q.I\rvert,\;k\bigr)$
    \STATE \quad $\mathit{output} \leftarrow$ first $\mathit{cnt}$ elements of $Q.I$, removing them from $Q.I$
    \STATE \quad \textbf{return} $\mathit{output}$
\end{algorithmic}
\end{algorithm}

\begin{algorithm}[t!]
\caption{Incremental search function}
\label{alg:incremental-search}
\begin{algorithmic}[1]
    \REQUIRE 
        $q\_id$, 
        $k$, 
        $b$, 
        $mx\_inc$, 
        $E$ : See Algorithm~\ref{alg:new-search} for descriptions
    \ENSURE 
        $q\_id$ is valid and updates priority queue $T$ and item list $I$ inside QS[$q\_id$].
        
    \STATE \textbf{function} \textsc {IncrementalSearch}($q\_id,\,k,\,b,\,mx\_inc,\,E$)
    \STATE \quad Let $Q \leftarrow$ $QS[q\_id]$,\; 
              $T \leftarrow Q.T$,\; 
              $I \leftarrow Q.I$, \; 
              $leaf\_cnt \leftarrow 0$, \; 
              $increments \leftarrow 0$
    \STATE \quad \textbf{if} $T$ is empty \textbf{then}
    \STATE \quad\quad $root\_distances \leftarrow \mathrm{calculate\_distances}(\texttt{root},\,Q.q,\,\texttt{metric})$
    \STATE \quad\quad Let $leaf \leftarrow \bigl(L = 1\bigr)\;?\;1\;:\;0,\;level \leftarrow 0$
    \STATE \quad\quad \textbf{for each} root child $c$ with distance $d$:
    \STATE \quad\quad\quad Push into $T$: $(\,d,\;leaf\_flag,\;level,\,c\,)$
    
    \STATE \quad \textbf{while} $T \neq \emptyset$ \textbf{do} ~~~~~~~~~~~~~~~~~~~~~~~~~~~~~~~~~~~~~~~~// Main tree‐search loop
    \STATE \quad\quad Pop one entry $(is\_leaf,\,level,\,node)$ from $T$
    \STATE \quad\quad Retrieve $embeddings$ for $(level,\,node)$; \textbf{continue} if none
    \STATE \quad\quad Compute $distances \leftarrow \mathrm{calculate\_distances}(embeddings,\,Q.q,\,\texttt{metric})$
    \STATE \quad\quad \textbf{if} $is\_leaf$ \textbf{then} ~~~~~~~~~~~~~~~~~~~~~~~~~~~~~~~~~~~~// Cluster: Scan
    \STATE \quad\quad\quad Let $children \leftarrow$ children of $(level,\,node)$
    \STATE \quad\quad\quad \textbf{for each} child $c$ with distance $d$:
    \STATE \quad\quad\quad\quad \textbf{if} $c \notin exclude$ \textbf{then}
    \STATE \quad\quad\quad\quad\quad Add $(d,c)$ to $I$
    \STATE \quad\quad\quad\quad\quad $leaf\_cnt \leftarrow leaf\_cnt + 1$
    \STATE \quad\quad \textbf{else} ~~~~~~~~~~~~~~~~~~~~~~~~~~~~~~~~~~~~~~~~~~~~~~~~~~~~// Internal Node: Explore
    \STATE \quad\quad\quad Let $children \leftarrow$ children of $(level,\,node)$
    \STATE \quad\quad\quad \textbf{for each} child $c$ with distance $d$:
    \STATE \quad\quad\quad\quad $next\_is\_leaf \leftarrow (level+1 = \texttt{levels}-1)\,$? 1 : 0
    \STATE \quad\quad\quad\quad Push into $T$: $(\, d,\, next\_is\_leaf,\, level+1,\, c)$
    
    \STATE \quad\quad \textbf{if} $leaf\_cnt = b$ \textbf{then} ~~~~~~~~~~~~~~~~~~~~~~~~~~~~~// Loop break condition
    \STATE \quad\quad\quad \textbf{if} $|I| \ge k$ \textbf{then}
    \STATE \quad\quad\quad\quad Sort $I$ by score; \textbf{break}
    \STATE \quad\quad\quad \textbf{else if} $increments > mx\_inc$ \OR $mx\_inc = -1$ \textbf{then}
    \STATE \quad\quad\quad\quad $increments \leftarrow increments + 1$
    \STATE \quad\quad\quad\quad $b \leftarrow 2b$
    \STATE \quad\quad\quad \textbf{else}
    \STATE \quad\quad\quad\quad \textbf{break}
\end{algorithmic}
\end{algorithm}

Algorithm~\ref{alg:new-search} shows the $\mathrm{NewSearch}$ function that is used when a query is initially submitted. A new query state object is created and appended to the list of query states $QS$ and the query is assigned a query id ($|QS| - 1$). The query state object consists of the query $Q$, a priority queue $T$ and an items list $I$. Once the new query state has been added, the $\mathrm{IncrementalSearch}$ function is called to process $T$ and populate $I$, whereafter $\mathrm{GetNextKItems}$ is called to return the top $k$ items from $I$. The initial search function outputs both the results and the query id.
Algorithm~\ref{alg:get-next-k-items} shows the $\mathrm{GetNextKItems}$ function which can be called with the query id to get $k$ more items from that query. In case the $I$ for that query has less than $k$ items it calls $\mathrm{IncrementalSearch}$ to populate $I$ with more items. Algorithm~\ref{alg:incremental-search} shows the details of the incremental search approach using the priority queue.
If it is an initial search the \texttt{index\_root} embeddings are used to determine the nodes to open from \texttt{lvl\_1}.
Whenever a leaf node is accessed, its children are added to a vector $I$. Once $b$ leaf nodes (clusters) have been accessed, there is a check to see whether $k$ items have been found. 
In case there are $k$ items it returns them, otherwise, the algorithm doubles $b$ and continues going through the $T$. 

There is a potential pitfall in this new method as it is no longer upper-bound to explore only $b$ internal nodes per level. It is free to explore any number of internal nodes in the search for the $b$ best clusters and the $k$ most similar items. 
We could enforce the old behavior by having one $T$-queue per index level, capping each to only $b$ checks, but we opted to not implement this in the current version. 

\subsection{Implementation}
The described index has been implemented both in Python and in Rust (as a Rust-Python extension). 
Aside from the compiled-versus-interpreted distinction, the sole substantive difference is how node embeddings are loaded. The Python version keeps their original data type, whereas the Rust version converts them to \texttt{float32}.
Note that all the evaluations show results from the Rust version.
\section{Experimental Setup: MMIR Benchmark}

The Lifelog Search Challenge~(LSC)~\cite{lsc24} and Video Browser Showdown~(VBS)~\cite{lokovc2023interactive} are fast-paced, live competitions in which research teams have only minutes to solve retrieval tasks over large image- and video-based collections. The LSC and VBS categorize their tasks into three groups; known item search (KIS), ad-hoc search (AS), and question-answering (QA). The objective of KIS tasks is to find one relevant item from a tiny groundtruth set, and the task can be either textual (T-KIS) or visual (V-KIS). In T-KIS, a gradual step-by-step textual description is presented where each step adds more contextual or visual information towards the items to find. In V-KIS, a video clip is presented and the goal is to find any shot within that clip. The AS tasks also presents a textual description, but  the objective in this task is to find as many items matching that description as possible. The submitted items are judged live by a panel. Lastly the QA tasks pose a question around items from the collection and the objective is to find and analyze the items and submit a textual answer also judged live by the panel. To solve these tasks, teams draw from a multitude of components rather than a fixed recipe: they might use high-dimensional indexes for text, video-level or key-frame embeddings, plug in metadata databases for instant keyword look-ups, or call on foundation models---LLMs for query reformulation, VLMs for captioning or event localization---whenever resources allow. Each chosen piece consumes compute and memory, so strategies diverge: some off-load heavy steps to the cloud (gambling on network stability), while others compress or reduce data representations and accept a modest accuracy hit~\cite{amato2024visione,nguyen2024eagle,sharma2025exquisitor,hezel2024libro}.
To explore these trade-offs in a controlled setting, we constructed an experimental benchmark that replays the LSC and VBS (textual known-item-search) tasks on their datasets.
Because all queries and data are from the challenges, the benchmark preserves the realistic time pressure and content diversity they impose. Under these constraints, structures such as the eCP-FS index can better show their value, providing competitive retrieval speed while staying within the tight hardware budgets typical of live events with local machines.
The benchmark is implemented in an extendable manner to allow extending task types and workloads in the future.

\subsection{Datasets}

The benchmark consists of 3 datasets, the Lifelog Search Challenge 2024~\cite{lsc24} dataset~(LSC24), the V3C1 and V3C dataset from Video Browser Showdown~\cite{v3c}. Their details are listed in Table~\ref{tab:mmir:datasets}. We have extracted embeddings from each dataset using a vision-language model that utilizes a Sigmoid loss for Language-Image Pre-training (SigLIP) model trained on the WebLI dataset and provided via OpenCLIP~\cite{zhai2023sigmoid,chen2023pali,IlharcoGabriel2021}. The resulting embedding dimension is 1152 and stored in \texttt{float16}.

\begin{table}[t!]
    \centering
    \caption{Dataset information}
    \begin{tabular}{l@{~~~~}l@{~~~~}r@{~~~~}c}
        \toprule
        \bf Dataset & \bf Media Type & \multicolumn{1}{c}{\bf Items} & \bf Size~(GB) \\
        \midrule
        LSC24 & Images & 725,226 & 1.43 \\
        V3C1 & Videos (Keyframes) & 1,007,360 &  1.97 \\
        V3C & Videos (Keyframes) & 4,143,681 &  8.93 \\
        \bottomrule \\
    \end{tabular}
    \label{tab:mmir:datasets}
\end{table}

\subsection{Tasks \& Workloads}
The tasks we  use for the benchmark is textual known item search (T-KIS). In the live setting users typically create their own queries that relate with their systems underlying model(s) representation. 
In VBS, the task descriptions are presented in 3 steps and in LSC in 6 steps. For the purpose of this evaluation, we use all the descriptions at any step as individual queries.  The tasks for VBS have been collected from the 2019-2024 editions, and for LSC it is from 2022-2024, leading to 63 tasks (189 queries) and 34 tasks (204 queries), respectively.

The evaluation will use two workloads; the first evaluates the single query performance, while the other evaluates the incremental query performance. 
The single query performance is straightforward, where each query is run with a request to retrieve the top 100 items. Each query is run 10 times to get a realistic average query latency. For disk-based indexes, the first run latency is categorized as disk, while the remainder is in-memory.
The incremental query workload starts by searching for the top 100 items of a given query and follows it up with requesting 100 more items 10 times. Each query is run 10 times in this manner.

We run the benchmark on a laptop with the following specifications: Windows 11 Home, 16 GB RAM (4800 MHz), SSD (NVMe Micron 3400), Intel i9-12900H (2.5 GHz, 14 cores, 20 logical processes).
We compare eCP-FS against one other disk based system, DiskANN, and two in-memory systems, IVF and HNSW~\cite{douze2024faiss}.
The eCP-FS indexes have been build using $L=3$ for V3C and $L=2$ for V3C1 and LSC24. 
The target cluster size $C$ is 455 vector embeddings and with 1152 \texttt{float16} dimensions per vector that is roughly 1 MB of data per cluster.
For the cluster-based systems we set the search expansion, eCP's $b$ and IVF's \texttt{nprobe}, to 64.
For the graph-based indexes we use $k=100$ as their search complexity parameter (\texttt{efSearch}  for HNSW and \texttt{complexity} for DiskANN).
Since IVF and HNSW cannot use V3C with all dimensions in float32 (exceeds the 16 GB RAM limit), we have used PCA to reduce the dimensions to 720 (0.99 EVR) and 496 (0.95 EVR). 
The current available DiskANN implementation requires that the full data set be in memory to build it, so it also uses the PCA versions for V3C. 

The quality of results is not the goal of this evaluation, but rather the latency hit for using a file structure based index. In general, eCP is  on par or slightly lower in accuracy than the other indexes, but provides better scalability.

\section{Results and Analysis / (Experimental) Evaluation}

The aim of these experiments is to analyze the behavior of eCP-FS with search-oriented workloads, and see what benefits the file structure provides over the optimized in-memory indexes IVF and HNSW, and less flexible disk based index of DiskANN.

\begin{table}[t!]
    \centering
    \caption{Index load times, average query latencies, and average workload latency in seconds of single query workload over 10 runs. The () in the eCP-FS Workload column refers to the in-memory workload average.}
    \small
    \begin{tabular}{p{2.5cm}p{2cm}cccc}
        \toprule
        &  & &\multicolumn{3}{c}{\bf Latency}\\
        \cmidrule{4-6}
        \bf Collection & \bf Index & \bf Load Time~~ & \bf Disk~~ & \bf Memory~~ & \bf Workload\\
        \midrule
        \bf LSC24 & IVF & 1.152 & - & 0.005 & 1.168 \\
         & HNSW & 3.328 & - & 0.0004 & 0.082 \\
         & DiskANN & 79.780 & 0.014 & - & 2.831 \\
         & eCP-FS & 0.38 & 0.260 & 0.009 & 6.994 (1.869) \\
        \midrule
        \bf V3C1 & IVF & 2.967 & - & 0.011 & 1.212 \\
         & HNSW & 4.520 & - & 0.001 & 0.110 \\
         & DiskANN & 102.144 & 0.021 & - & 2.352 \\
         & eCP-FS & 0.86 & 0.729 & 0.023 & 10.428 (2.588) \\
        \midrule
        \bf V3C  & IVF & 7.366 & - & 0.003 & 0.612 \\
        \bf (496) & HNSW & 7.998 & - & 0.0016 & 0.304 \\
         & DisKANN & 88.994 & 0.011 & - & 2.094 \\
         & eCP-FS & 3.581 & 0.452 & 0.008 & 10.0 (1.615) \\
        \midrule
       \bf  V3C  & IVF & 11.355 & - & 0.005 & 0.955 \\
         \bf (720)& HNSW & 18.866 & - & 0.002 & 0.426 \\
         & DiskANN & 96.803 & 0.011 & - & 2.198 \\
         & eCP-FS & 3.365 & 0.619 & 0.014 & 14.112 (2.663) \\
       \midrule
       \bf  V3C  & IVF & - & - & - & - \\
\bf (1152)         & HNSW & - & - & - & - \\
         & DiskANN & - & - & - & - \\
         & eCP-FS & 3.803 & 1.267 & 0.016 & 26.7 (3.049) \\
        \bottomrule \\
    \end{tabular}
    \label{tab:exp1:latency}
\end{table}

\subsection{Single Query Performance} 
Table~\ref{tab:exp1:latency} shows the results from the  average latencies for each collection and index for the single query workload. It comes as no surprise that the IVF and HNSW indexes are the fastest, since they are heavily optimized for in-memory performance. However, once eCP-FS has loaded its nodes into memory, subsequent searches for the same queries or queries falling into similar paths are only slower by roughly a factor of 2 from IVF. Looking at the disk average query latency eCP-FS is considerably slower than DiskANN, which is also expected due to the overhead of opening and closing thousands of files in the file structure. In contrast, DiskANN works with a small set of  serialized files, leading to far more controlled access. To emphasize the difference between disk and in-memory performance of eCP-FS, we examine the first workload which holds no node data in memory. For LSC24 with 204 queries per workload this leads to $0.26 \times 204 = 53$ seconds, whereas the following 9 workloads take on average  $0.009 \times 204 = 1.8$ seconds each.
\begin{table}[t]
    \centering
    \caption{Tasks completed in each collection, meaning that 1 relevant item from the ground truth was in the top 100 returned items.}
    \begin{tabular}{lccccc}
        \toprule
         \bf Index & \bf LSC24 & \bf V3C1  & \multicolumn{3}{c}{\bf V3C}\\
         \cmidrule{4-6}
       &&& \bf (496) & \bf (720) & \bf (1152) \\
        \midrule
        IVF & 31/34 & 29/37 & 26/63 & 14/63 & - \\
        HNSW & 30/34 & 31/37 & 10/63 & ~~1/63 & - \\
        DiskANN & 31/34 & 30/37 & 33/63 & ~~1/63 & - \\
        eCP & 30/34 & 21/37 & ~~8/63 & ~~4/63 & 27/63 \\
        \bottomrule \\
    \end{tabular}
    \label{tab:tasks}
\end{table}

While we have stated quality is not of particular interest in this evaluation, Table~\ref{tab:tasks} shows the number of tasks each index managed to solve. Recall that each task has 6 queries in LSC and 3 queries in V3C1 and V3C. If one of the queries manage to get an item from the ground truth in the top 100, that task is considered complete. eCP-FS performs as expected, capable of solving tasks but not as effectively as the other indexes. An interesting observation is the V3C tasks, where PCA 496 is better than PCA 720, and how much eCP-FS benefits from being able to use the standard embeddings.

In terms of memory footprint, the HNSW and IVF index both load the entire dataset, while DiskANN loads a considerably small graph along with using PQ compressed vectors further minimizing memory. eCP-FS does not load the entire dataset and does not utilize any form of compression to reduce memory when data is loaded. eCP-FS stores the data in its original format, i.e. in this case \texttt{float16}, but loads it into \texttt{float32}. Therefore, it is capable of exceeding available memory if the loaded data is not kept in check through an LRU style caching mechanism. It is also possible to set caching off, such that it frees the nodes memory after use. In practice, however, single user sessions may not  involve enough diverse queries to load significant portions of a large dataset.

\subsection{Incremental Search Performance}
Table~\ref{tab:exp2:latency} shows the average query latencies from the incremental search workloads, where each query is run once with $k=100$ followed by 10 requests for $k$ more items. For the eCP-FS index, any follow-up request for a query results in checking its query state object $Q$ for the requested query. If $Q.I$ contains $k$ or more items it is only a matter of returning them, and in case there is not enough it resumes the search using the priority queue $Q.T$. As the other indexes do not have an internal incremental search implementation, nor do they keep state of queries, the request for additional $k$ results in requesting $k + (k \times rd)$, where $rd$ is the number of rounds going from 0 to 9. Due to this implementation detail it is not surprising that eCP-FS outperforms the other indexes in this workload. 
Note that while the disk latency is again slower than DiskANN's, it is a one time cost per query, whereafter it is primarily just returning items from $Q.I$ in memory. 
This leads to a staggering difference in average workload times in eCP-FS favor. Similar to the first workload, eCP-FS has to pay the cost of loading from disk one time (as no upper node in memory limit was set). 
While the other indexing methods could potentially be extended to support similar incremental query states, this adaptation would inevitably increase their memory requirements. Specifically, these indexes would require additional memory to maintain query state information, in addition to the already significant memory usage of the fully loaded indexes themselves for IVF and HNSW. 
In contrast, with eCP-FS it is possible to release the memory occupied by the index data itself while retaining only the necessary query states in memory. 
Furthermore, eCP-FS's underlying file structure can be extended to persistently store these query states. Such persistence provides an important advantage: it allows users to load only the minimal set of data (such as the specific item arrays associated with a query) initially, delaying loading additional structures like the priority queue until explicitly needed (e.g., when the current query results no longer meet the desired criteria).

\begin{table}[t!]
    \centering
    \caption{Average query latencies and average workload time in seconds for incremental search workload over 10 runs. The () in the eCP-FS Workload column refers to the in-memory workload average.}
    \small
    \begin{tabular}{p{2.5cm}p{2cm}cccc}
        \toprule
        & &\multicolumn{3}{c}{\bf Latency}\\
        \cmidrule{3-5}
        \bf Collection & \bf Index & \bf Disk~~ & \bf Memory~~ & \bf Workload\\
        \midrule
        \bf LSC24 & IVF & - & 0.002 & 34.654 \\
         & HNSW & - & 0.0004 & 8.373 \\
         & DiskANN & 0.021 & - & 428.459 \\
         & eCP-FS & 0.239 & 0.0001 & 6.636 (1.944) \\
        \midrule
        \bf V3C1 & IVF & - & 0.003 & 38.950 \\
         & HNSW & - & 0.002 & 24.148 \\
         & DiskANN & 0.028 & - & 315.856 \\
         & eCP-FS & 0.663 & 0.0002 & 9.574 (2.452) \\
        \midrule
        \bf V3C & IVF & - & 0.0011 & 20.893 \\
        \bf (496) & HNSW & - & 0.0011 & 21.694 \\
         & DiskANN & 0.014 & 0.014 & 266.206 \\
         & eCP-FS & 0.448 & 0.00008 & 9.867 (1.547) \\
        \midrule
        \bf V3C & IVF & - & 0.0016 & 31.100 \\
        \bf (720) & HNSW & - & 0.0010 & 19.055 \\
         & DiskANN & 0.013 & - & 247.558 \\
         & eCP-FS & 0.594 & 0.0001 & 13.328 (2.315) \\
        \midrule
        \bf V3C & IVF & - & - & - \\
        \bf (1152) & HNSW & - & - & - \\
         & DiskANN & - & - & - \\
         & eCP-FS & 1.218 & 0.0001 & 25.175 (2.378) \\
        \bottomrule \\
    \end{tabular}
    \label{tab:exp2:latency}
\end{table}



eCP-FS’s lean memory footprint and plug-and-play architecture make it ideal for long-running analytics, where the same query can remain active or be revisited over hours or days. Furthermore, it allows for multiple indexes to coexist, and gives fine-grained control over what query states or node data stay cached. The first response to a cold query is slightly slower, as our benchmarks show, but the savings in memory and the added flexibility far outweigh that one-off latency when overall resource efficiency matters more than raw, single-query speed. An additional discovery from exploring this type of hierarchical, file-based ANN index, we found that searches can run in parallel with construction. Early results are rough, but they improve steadily as the index fills with more data.


\section{Conclusion}

In this paper we have explored the potential of using a file structure as the basis of a hierarchical ANN index. Specifically, the paper presented eCP-FS, a disk-based ANN algorithm that stores both data and index structure on disk. 
Exposing the full index structure as a file system overcomes a common problem with state-of-the-art indexing algorithms, namely that valuable information is often hidden away either inside a black-box or in open source code that is highly optimized and very complex. 
As expected, eCP-FS I/O access costs are high. To alleviate this we propose a cache-like mechanism that allows us fine control over the memory footprint (limiting clusters in RAM + LRU policy) while still maximizing search efficiency using prefetching.
We also presented a highly efficient resume-search functionality that is of great value to modern search engines. The resume feature is achieved by retaining the search state after returning results, allowing follow up queries to be easily continued. This search features also takes full advantage of the caching / pre-fetching capabilities of eCP-FS.
We compared eCP-FS to one other disk-based algorithm, DiskANN, as well as two other in-memory algorithm HNSW and IVF. eCP-FS does indeed have the highest I/O costs in a cold state but once the cache is hot it is not far off the other disk-based algorithm.
Where eCP-FS shines is in the resume-search scenarios. This is to be expected as the others do not support this functionality, i.e.~they start each search from scratch.
The experiments do however highlight the advantages of this functionality. 
Finally we claim that eCP-FS offers great flexibility and adaptability, making it ideal for dynamic retrieval tasks where resources are scarce. While outperformed by specialized and optimized in-memory systems (IVF and HNSW), eCP-FS does have the incremental retrieval capability, something pure search-oriented systems typically lack. eCP-FS also offers greater versatility compared to rigid disk-based solutions like DiskANN.

\begin{credits}
\subsubsection{\ackname} This work was supported by Icelandic Research Fund grant 239772-051.

\end{credits}

\bibliographystyle{splncs04}
\bibliography{references}

\end{document}